\documentclass[aps,prb,reprint,superscriptaddress]{revtex4-2}

\usepackage{amsmath}
\usepackage{amssymb}
\usepackage{xcolor}

\usepackage{amsmath,amssymb,amsfonts,bm,color,graphicx}

\usepackage{tabularx,multirow,array,diagbox}

\usepackage{bm}

\usepackage[unicode=true,colorlinks=true]{hyperref}

\usepackage[etex=true,export]{adjustbox}

\usepackage{ulem}
\newcommand{\beq}{\begin{equation}}
\newcommand{\eeq}{\end{equation}}
\newcommand{\beqn}{\begin{eqnarray}}
\newcommand{\eeqn}{\end{eqnarray}}

\usepackage{graphicx} 

\begin{document}
\title{Phase-sensitive non-reciprocal transport in high-temperature superconductor}

\author{Guo-Liang Guo}
\email{sjtu2459870@sjtu.edu.cn}
\affiliation{Tsung-Dao Lee Institute, Shanghai Jiao Tong University, Shanghai 200240, China}

\author{Xin Liu}
\email{phyliuxin@sjtu.edu.cn}
\affiliation{Tsung-Dao Lee Institute, Shanghai Jiao Tong University, Shanghai 200240, China}
\affiliation{Hefei National Laboratory, Hefei 230088, China National Laboratory, Hefei 230088, China}
\affiliation{Shanghai Research Center for Quantum Sciences, Shanghai 201315, China}

\begin{abstract}

We propose the superconducting diode effect (SDE) in a planar s–wave/d–wave/s–wave Josephson junction as a direct phase‑sensitive probe of the d‑wave pairing function in high‑$T_c$ superconductors. Asymmetric interface coupling breaks inversion symmetry and induces a spontaneous $\pm\pi/2$ phase difference, thereby breaking time‑reversal symmetry without a magnetic field. In this TRS‑broken state, the SDE emerges when single-Cooper-pair tunneling is enabled at the s-d interfaces, with its polarity and efficiency controllable by rotating the d‑wave crystallographic orientation or perturbing its intrinsic $C_4$ symmetry. Our results reveal a robust link between nonreciprocal Josephson transport and pairing symmetry, establishing the SDE as a powerful diagnostic tool for high‑$T_c$ superconductors and a tunable element for superconducting electronics.

\end{abstract}

\maketitle

{\it Introduction $-$} Nonreciprocal transport, a hallmark of semiconductor p-n junctions \cite{Scaff1947,Shockley1949}, also manifests in superconducting systems as a non-dissipative directional asymmetry in the critical supercurrent—a phenomenon known as the superconducting diode effect (SDE). This remarkable effect has garnered considerable attention in recent years \cite{Jiang2022,Zhang2022,Nadeem2023,Kochan2023,Mei2025,Wang2025}, due to its unique charge transport properties and potential applications in next-generation cryoelectronic devices. In such systems, the magnitude of the critical current differs for opposite directions ($I_{c+}\neq I_{c-}$), enabling dissipationless current flow in one direction while being resistive in the other. This non-reciprocity effect has also been observed and explored in various systems involving Josephson junctions \cite{Pal2022,Davydova2022,Pal2022,Steiner2023,Maiani2023,Steiner2023}, commonly referred to as the Josephson diode effect (JDE). In general, a nonreciprocal supercurrent requires the simultaneous breaking of time-reversal symmetry (TRS) and inversion symmetry, and is typically accompanied by a non-sinusoidal current–phase relation (CPR), where higher Fourier components play a critical role \cite{Yuan2022}. The TRS breaking can arise from the magnetic or exchange fields \cite{Legg2022,Narita2022,Legg2023,Lu2023}, for which the JDE polarity is often an odd function of the applied field. More recently, field-free JDE has been demonstrated in several platforms, pointing to mechanisms based on spontaneous TRS breaking \cite{Lin2022,Hu2023,Wu2022}. 

Particularly, the field-free JDE in high-temperature (high-$T_c$) superconductors with spontaneous TRS breaking is widely explored due to their high critical temperature and unconventional pairing function \cite{Hirschfeld1993,McElroy2005,Leng2012,Fradkin2015,Cai2016,Jacobs2016,Liao2018}, both advantageous for further applications and the generation of JDE \cite{Tummuru2022,Zhu2023,Zhao2023,Volkov2024,Ghosh2024,Wang2025a}. Despite extensive experimental evidence from angle-resolved photoemission spectroscopy, scanning tunneling microscopy, and phase-sensitive measurements supporting predominant d-wave pairing in cuprates, the pairing symmetry remains under debate \cite{Tsuei2000,Zhu2021,Wang2023,Wang2025a}. Growing experiments and theoretical analysis suggest a possible coexistence of an isotropic $s$-wave component and $C_4$ symmetry $d$-waves superconducting gap in cuprates \cite{Zheng2025,Wang2025a}, which directly impacts Josephson coupling and, intriguingly, is often associated with finite JDE \cite{Zhu2023,Zhao2023,Volkov2024,Ghosh2024,Wang2025a}. These observations highlight the opportunity to use nonreciprocal supercurrent not merely as an engineering functionality, but as a phase-sensitive probe of the superconducting order parameter in unconventional superconductors.

\begin{figure}[!htbp]
	\centering
	\includegraphics[width=1\columnwidth]{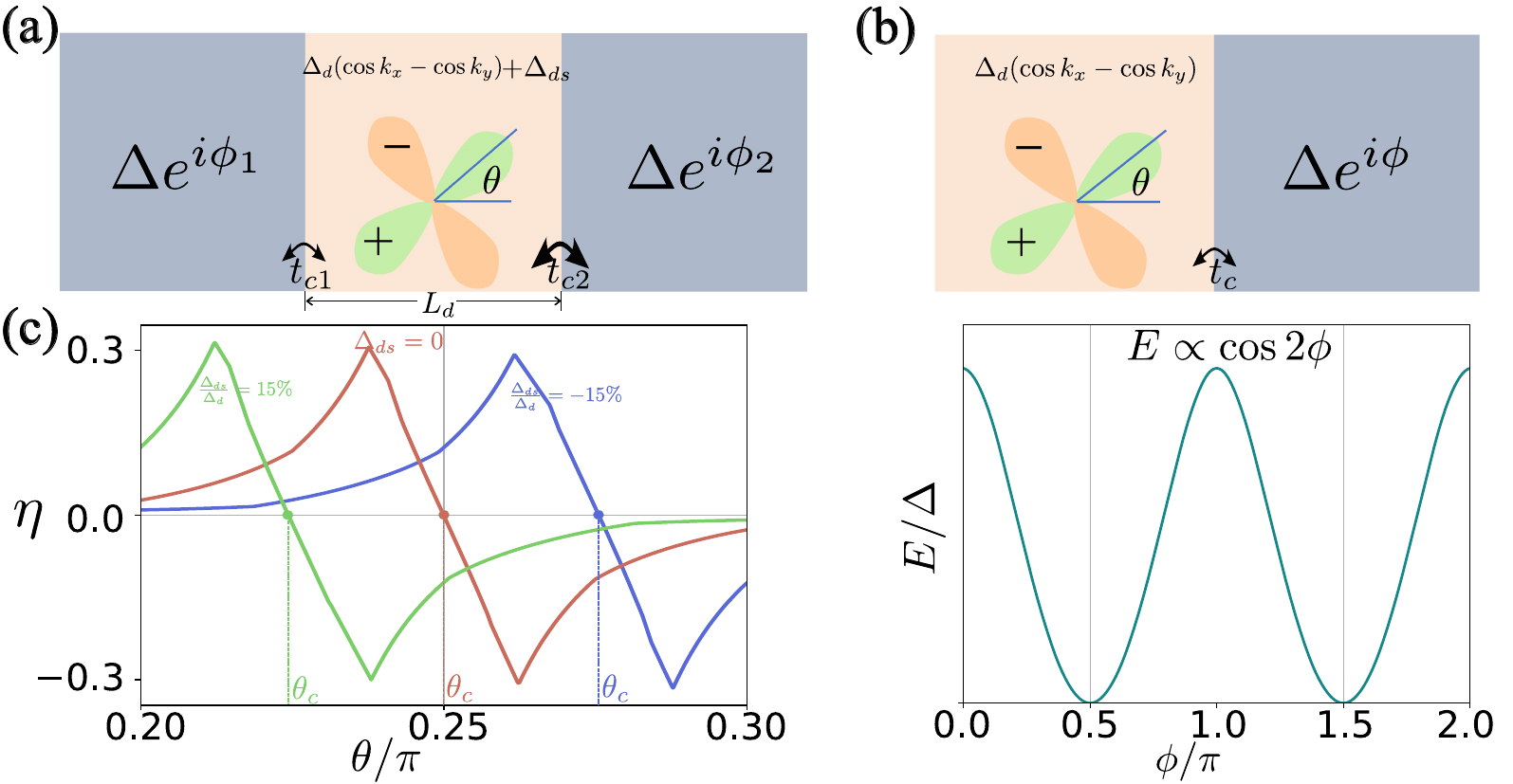}
	\caption{(a) Schematic diagram of planar s-d-s Josephson junction with asymmetric s-d interface couplings $t_{c1}<t_{c2}$, $\theta$ the lobe angle of $d$-wave pairing function, $L_d$ the $d$-wave length. (b) Schematic diagram of s-d Josephson junction, with the potential $E\propto\cos2\phi$ at $\theta=\pi/4,\Delta_{ds=0}$. (c) Diode efficiency as a function of $\theta$ with different $\Delta_{ds}$, with $\theta_c$ the critical point for reversing the polarity of $\eta$.
  }
\label{dev}
\end{figure}

In this work, we exploit this concept by proposing a planar $s$–$d$–$s$ Josephson junction [Fig.\ref{dev}(a)] as a field-free, symmetry-based probe of $d$-wave pairing. By intentionally engineering asymmetric $s$–$d$ interface couplings, $t_{c2}>t_{c1}$, inversion symmetry is broken. For a d-wave lobe angle $\theta=\pi/4$, the s–d coupling forbids single–Cooper-pair tunneling, yielding a dominant second harmonic, $E(\phi)\propto\cos(2\phi)$, with minima at $\phi=\pm\pi/2$ [Fig.~\ref{dev}(b)]. This spontaneously breaks TRS by pinning a $\pm\pi/2$ phase difference at the stronger interface in planar $s$–$d$–$s$ junction. However, due to a residual $\mathbb{Z}_2$ symmetry, $E(\Phi+\phi)=E(\Phi-\phi)$, the junction behaves as a $\phi_0$ junction without exhibiting the JDE. A finite JDE becomes feasible when single–Cooper-pair tunneling is symmetry-allowed at the $s$–$d$ interface. This can be realized by rotating the $d$-wave crystallographic orientation away from $\pi/4$,  or by breaking its intrinsic $C_4$ pairing symmetry. The maximal diode efficiency $\eta\equiv(I_{c+}-|I_{c-}|)/(I_{c+}+|I_{c-}|)$ can approach 1/3, and its polarity reverses across a critical angle $\theta_c$, with $\theta_c = \pi/4$ for $\Delta_{ds} = 0$ and $\theta_c<(>)\pi/4$ for $\Delta_{ds}>(<)0$ [Fig.~\ref{dev}(c)]. Our findings establish a direct and controllable link between JDE polarity/efficiency and the $d$-wave gap structure in high-$T_c$ superconductors, providing both a practical route to implement field-free SDE and a powerful phase-sensitive diagnostic of pairing symmetry—without the need for an external magnetic field.

The Ginzburg-Landau (GL) free energy of the system in Fig.~\ref{dev}(a) can be written as
\begin{equation}
\begin{aligned}
    \mathcal{F}[\psi_{s1},&\psi_{s2},\psi_d]=\mathcal{F}_0[\psi_{s1}]+\mathcal{F}_0[\psi_{s2}]+\mathcal{F}_0[\psi_{d}]\\
    &+\sum_{i=1,2}B_i(\psi_{si}^*\psi_d+c.c.)+C_i(\psi_{si}^{*2}\psi_d^2+c.c.)\\
    &+D(\psi_{s1}^*\psi_{s2}+c.c.),
\end{aligned}
\end{equation}
where complex scalars $\psi_{a}(a=s1,s2,d)$ are the order parameters, and $|\psi_a|\sim\sqrt{1-T/T_{c,a}}$ near there critical temperature and usually $T_{c,s}\ll T_{c,d}$. $\mathcal{F}_0[a]$ donate GL free energies of the individual superconductors. Physically, the coefficients $B_{1(2)}$ and $C_{1(2)}$ terms represent coherent tunneling of single and double Cooper pairs between
the $s$-wave and $d$-wave superconductors, while $D$ the effective coupling between the two $s$-wave superconductors. If the $d$-wave superconductors obey tetragonal symmetry, the coefficient $B_{1(2)}$ is required to vanish when the lobe angle $\theta=\pi/4$ \cite{Xiang2022}. With the phase difference of the two $s$-wave superconductors relative to the $d$-wave superconductor $\phi_{1(2)}$, the free energy becomes
\begin{equation}
\begin{aligned}
        \mathcal{F}&(\phi_1,\phi_2)=\mathcal{F}_0+2C_1|\psi_{s1}|^2|\psi_{d}|^2\cos2\phi_1\\
  &+2C_2|\psi_{s2}|^2|\psi_{d}|^2\cos2\phi_2+2D|\psi_{s1}||\psi_{d}|\cos(\phi_1-\phi_2),
  \label{GL-Fe}
\end{aligned}
\end{equation}
where $\mathcal{F}_0$ collects all terms independent of $\phi_{1(2)}$ and usually $C_{1,2}>0$ \cite{Tummuru2022,Patel2024,Volkov2024}. However, recent experiments indicate an $s$-wave pairing component in high-$T_c$ cuprates \cite{Tsuei2000,Zhu2021,Wang2023,Wang2025a}. The term $B_i$ will be nonzero even at $\theta=\pi/4$, and gives a conventional $2\pi$ periodic contribution to $\mathcal{F}(\phi_1,\phi_2)$ proportional to $\cos \phi_{1(2)}$.

{\it Model Hamiltonian.$-$} To capture the essential physics of the system, we then work with a continuum microscopic model Hamiltonian of the planar s-d-s Josephson junction (Fig.~\ref{dev}(a)) as $H=\sum_{i=1,2}(H_{si}+H_{Ti})+H_d$, where $H_{s1(s2,d)}$ the BCS model Hamiltonian of two $s$-wave and $d$-wave superconductors, respectively. $H_{T1(T2)}$ describe electron tunneling between the two $s$-wave and $d$-wave superconductors. The explicit forms are
\begin{equation}
\begin{aligned}
    H_{s1(s2,d)}&=\sum_{k(p,m)}\epsilon_{k}(c_k^\dagger c_k+c_{-k}^\dagger c_{-k})+\Delta^{(s,d)}(c_k^\dagger c_k^\dagger+c_{-k}^\dagger c_{-k}^\dagger)\\
    H_{T1}&=T_{kp}(c_k^\dagger c_p+h.c.)\\
    H_{T2}&=T_{mp}(c_m^\dagger c_p+h.c.),
    \label{sds-hat}
\end{aligned}
\end{equation}
where $k,p,m$ the momentum index in s-wave/d-wave/s-wave superconductors, respectively. $c_{k(m,p)}$ is the annihilation operator of electrons in the corresponding superconductors. $\epsilon_{k(m,p)}$ the kinetic energy, $\Delta^{(s)}=\Delta$ the $s$-wave superconducting gap and $\Delta^{(d)}=\Delta_d\cos2(\alpha-\theta)+\Delta_{ds}$ the $d$-wave superconducting gap with $\alpha$ the polar angle of the momentum, $\theta$ the angles between the x-axis and the lobe direction of the pairing function of the $d$-wave superconductor, as shown in Fig.~\ref{dev}(a). $\Delta_{ds}$ indicates the possible isotropic $s$-wave pairing function in the $d$-wave superconductor. The tunneling amplitudes depend on the momentum component perpendicular to the interface and are approximated as $T_{kp(pm)}=t_{c1(c2)}\cos\alpha$ \cite{Xiang2022}. The zero-temperature Josephson potential of the system is numerically obtained with the tight-binding (TB) based Hamiltonian \cite{supp}. 


\begin{figure*}[!htbp]
	\centering
	\includegraphics[width=2\columnwidth]{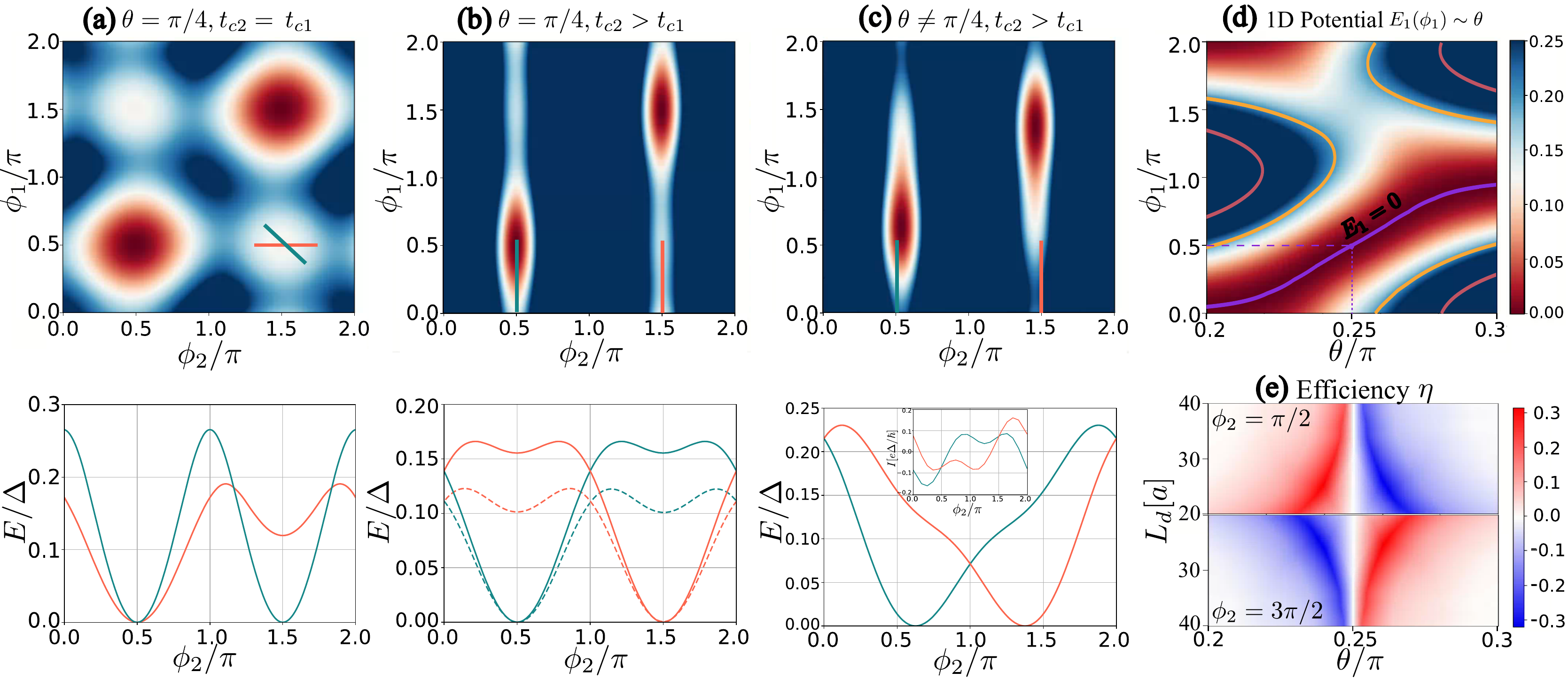}
	\caption{The 2D Josephson potential of s-d-s junction (normalized to 0.25) with the condition $\theta=\pi/4,t_{c2}=t_{c1}$ (a), $\theta=\pi/4,t_{c2}=4t_{c1}$ (b), $\theta=0.26\pi,t_{c2}=4t_{c1}$ (c), lower panels are the line-cut plot of the 2D potential, inset in (c) shows the current phase relation, (a)(b)(c) are calculated with $t_{c1}=0.25t, L_d=26a$, the dashed lines in lower panel (b) are calculated with $L_d=30a$. (d) With $\phi_2=\pi/2$, 1D Josephson potential $V_1(\phi_1)$ of s-d-s junction as a function of $\theta$, with $t_{c2}=4t_{c1}$. The solid lines are contour lines, the purple line is $V_1(\phi_1)=0$ (d) Diode efficiency as a function of $\theta$ and $d$-wave superconductor length $L_d$ with $\phi_2=\pi/2,3\pi/2$.
  }
\label{sds-jde}
\end{figure*}

{\it JDE in planar s-d-s junction.$-$} Since the Hamiltonian in Eq.~\eqref{sds-hat} respects TRS, the Josephson potential is an even function of the phases and contains only cosine terms. For analytical convenience, we first consider the case with inversion symmetry (symmetric s-d interface couplings) and the lobe angle $\theta=\pi/4$. In this scenario, the Josephson potential shown in Fig.~\ref{sds-jde}(a) is symmetric under exchanging $\phi_1$ and $\phi_2$ and has unequal valley along $\phi_{1(2)}=\pi/2$. The free energy can be approximated as
\begin{equation}
    V_1(\phi_1,\phi_2)=E_{J,1}^{(2)}\cos2\phi_1+E_{J,2}^{(2)}\cos2\phi_2+E_{J,s}^{(1)}\cos(\phi_1-\phi_2).
    \label{2d-sds-pot}
\end{equation}
where the first two terms correspond to two Cooper pair tunnelings across the s-d interfaces. With the Hamiltonian in Eq.~\eqref{sds-hat}, the coefficients $E_{J, i}^{(2)}(i=1,2)$ can be theoretically calculated and are proportional to $t_{c1}^4$ and $t_{c2}^4$ in the weak coupling regime \cite{supp}. The third term is the effective coupling between the two $s$-wave superconductors via the $d$-wave superconductor, with the coefficient $E_{J,s}^{(1)}$ proportional to $t_{c1}^2 t_{c2}^2$ \cite{supp}.

With asymmetric s-d interlayer couplings (here, we take $t_{c2}>t_{c1}$), the free energy develops deeper valleys near $\phi_2=\pi/2$ but shadowed along the $\phi_1$ direction (Fig.~\ref{sds-jde}(b)) and becomes non-invariant under the exchange of $\phi_1$ and $\phi_2$, consistent with the inversion symmetry breaking. These anisotropic valleys can lock the low-energy states and the phase difference $\phi_2$ to one of the deeper valleys (for concreteness, we take $\phi_2=\pi/2$), thereby spontaneously breaking TRS. Note that, with the GL free energy in Eq.~\eqref{GL-Fe}, the coefficients are related to the temperature as $(1-T/T_{c,s})$, the anisotropic valleys are valid in low temperature $T\ll T_{c,s}$. Then, the 2D potential simplifies to a 1D form as
\begin{equation}
    E(\phi_1)=E_{J,2}^{(2)}\cos2\phi_1\pm E_{J,s}^{(1)}\sin\phi_1,
    \label{sds-1D-C4}
\end{equation}
where "$\pm$" corresponds to $\phi_2$ located at $\pi/2$ or $3\pi/2$, and can be adjusted by external current bias exceeding the maximal current in the $\phi_2$ direction \cite{Volkov2024}. Then, the system becomes the $\phi_0$ junction and exhibits finite Josephson current at $\phi_1=0$, as shown in Fig.~\ref{sds-jde}(b). Notably, the $E_{J,s}^{(1)}\sin\phi_1$ is the Cooper pair tunneling between the two $s$-wave superconductors via the $d$-wave superconductor. Thus, increasing the length of $d$-wave superconductors will decrease the couplings between the two $s$-wave superconductors and increase the $\sin\phi_1$ component in the potential (dashed lines in Fig.~\ref{sds-jde}(b)). 

The asymmetric s-d interface couplings lock one of the phase differences to its minimal $\phi_2=\pi/2$, which breaks the inversion and TRS of the system, simultaneously. However, the Josephson potential in Eq.~\eqref{sds-1D-C4} exhibits reciprocal transport due to the residual $\mathbb{Z}_2$ symmetry, $E(\Phi+\phi_1)=E(\Phi-\phi)$ \cite{}, which yields equal forward and backward critical currents and, therefore, no JDE in this case (Fig.~\ref{sds-jde}(b)). This illustrates that a $\phi_0$ junction has no direct relationship with non-reciprocal transport, as it only relates to the maximum and minimum of the current and has nothing to do with the position where the current is zero \cite{Banerjee2023,Reinhardt2024,Guo2025}. 

The non-reciprocal transport becomes feasible when the lobe angle $\theta$ slightly deviates from $\pi/4$, which enables single Cooper pair tunneling across the s-d interfaces. In this scenario, the 2D Josephson potential of the planar s-d-s junction system (Fig.~\ref{sds-jde}(c)) maintains two deeper valleys along the $\phi_2$ direction, with the minimum remaining at $\phi_2\approx\pi/2$. This persistence arises because the larger interface coupling pins the minimum against small deviations of $\theta$ from $\pi/4$ (elaborated in the next section). Consequently, the phase difference $\phi_2$ remains locked at its minimum $\phi_2\approx\pi/2$, and the potential becomes the 1D form as
\begin{equation}
    E_1(\phi_1)=E_{J,1}^{(2)}\cos2\phi_1+E_{J,1}^{(1)}\cos\phi_1\pm E_{J,s}^{(1)}\sin\phi_1,
    \label{sds-1D-jde}
\end{equation}
as shown in Fig.~\ref{sds-jde}(c). The inset in Fig.~\ref{sds-jde}(c) clearly shows unequal positive and negative critical currents, indicating the emergence of the JDE. This is because the nonzero $\cos\phi_2$ term in the Josephson potential Eq.~\eqref{sds-1D-jde} breaks all $\mathbb{Z}_2$ symmetry and the JDE becomes feasible. Moreover, as the even number of Cooper pairs tunneling ($\cos2\phi_1$ term) is dominant with $\theta\approx\pi/4$, the breaking of all the $\mathbb{Z}_2$ symmetries requires the simultaneous presence of the $\cos\phi_1$ and $\sin\phi_1$ in the Josephson potential Eq.~\eqref{sds-1D-jde}.

Note that the coefficient $E_{J,1}^{(2)}$, which corresponds to double Cooper pair tunnelings, is always positive. The magnitude and polarity of the diode efficiency are completely determined by the magnitude and sign of the coefficients $E_{J,1}^{(1)}, E_{J,s}^{(1)}$, which depend on the angle $\theta$ and the minimal point of $\phi_2$. To demonstrate the effect of the angle $\theta$ deviating from $\pi/4$ on the Josephson potential, we thus set $\phi_2=\pi/2$ and calculate the 1D Josephson potential from Eq.~\eqref{sds-hat} while varying $\theta$ (Fig.~\ref{sds-jde}(d)). As $\theta$ deviates from $\pi/4$, the single Cooper pair tunneling is allowed, the potential minimum will shift from $\pi/2$ to $0$ or $\pi$ for $\theta>(<)\pi/4$ (the solid purple line), and the potential shape is symmetric with respect to $\theta=\pi/4$. Meanwhile, the term $E_{J,s}^{(1)}\sin\phi_2$ represents the effective coupling of the two $s$-wave superconductors via the $d$-wave superconductor; its amplitude is determined by the length of the $d$-wave superconductor $L_d$. We therefore calculate the diode efficiency as a function of $\theta$ and $L_d$ with fixed interlayer couplings and $\phi_2=\pi/2$, shown in Fig.~\ref{sds-jde}(e). The maximal $\eta$ approaches $1/3$, and its polarity reverses either when $\theta$ moves across $\pi/4$ or when $\phi_2$ switches between $\pi/2,3\pi/2$ \cite{supp}, consistent with changing the signs of the $\cos\phi_1$ or $\sin\phi_1$ terms in Eq.~\eqref{sds-1D-jde}.

{\it Spontaneous TRS breaking in s-d junction.$-$} Since strong interface coupling reinforces an “even-harmonic–dominated” energy landscape, even $\theta$ slightly deviates from $\pi/4$, which pins the phase near $\phi\approx\pi/2$ and robustly stabilizes spontaneous TRS breaking. This pinning mechanism provides a natural starting point for analysis. We thus quantify the dependence of the Josephson potential and the TRS-broken regime on interface coupling and $\theta$ in the minimal planar s–d junction, as shown in Fig.~\ref{dev}(b). The model Hamiltonian of the junction is $H=H_s+H_d+H_T$, where $H_T$ accounts for tunneling between the $s$- and $d$-wave superconductors. The explicit form of each term is
\begin{equation}
\begin{aligned}
    H_{s(d)}&=\sum_{k(p)}\epsilon_{k}(c_k^\dagger c_k+c_{-k}^\dagger c_{-k})+\Delta^{(s)(d)}(c_k^\dagger c_k^\dagger+c_{-k}^\dagger c_{-k}^\dagger),\\
    H_T&=T_{kp}(c_k^\dagger c_p+h.c.),
    \label{sd-hat}
\end{aligned}
\end{equation}
with $k,p$ the momentum index in s-wave/d-wave superconductors, respectively. The tunneling amplitude can be approximated as $T_{kp}= t_c\cos\alpha$ \cite{Xiang2022}. Since the Hamiltonian in Eq.~\eqref{sd-hat} respects TRS, the Josephson potential is an even function of phase difference $\phi$ and contains only cosine harmonics. Without loss of generality, retaining the leading harmonics, the Josephson potential can be written as
\begin{equation}
    V(\phi)=E_J^{(1)}(\theta)\cos\phi+E_J^{(2)}(\theta)\cos2\phi,
    \label{sd-pot}
\end{equation}

\begin{figure}[!htbp]
	\centering
	\includegraphics[width=1\columnwidth]{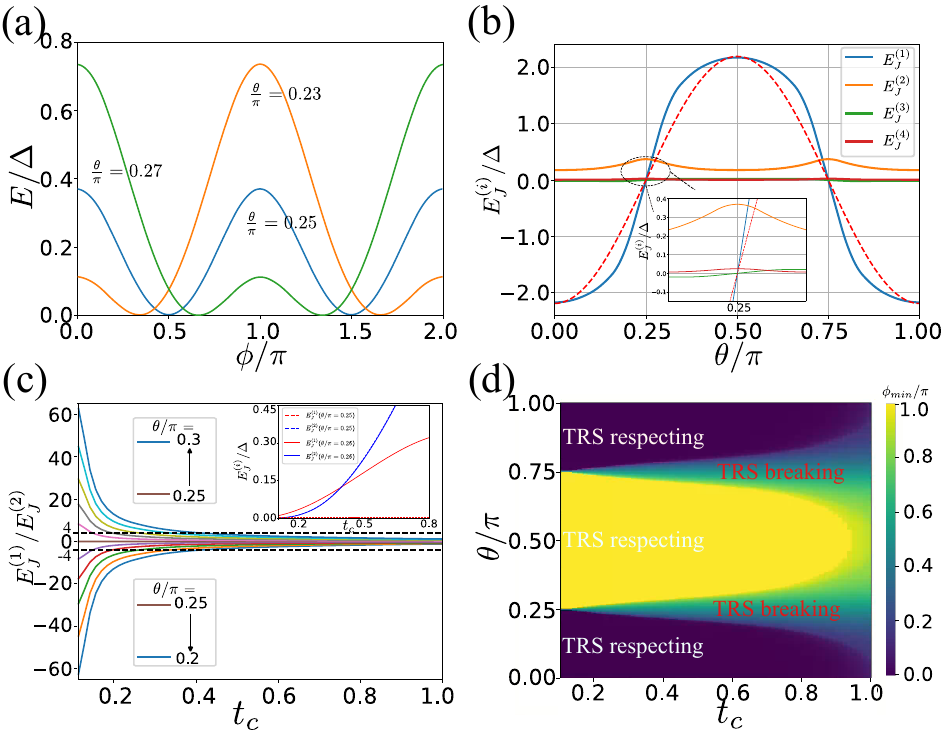}
	\caption{(a) The Josephson potential of the s-d junction with $\theta/\pi=0.23,0.25,0.27$. (b) The leading Fourier coefficients $E_J^{(i)}(i=1,2,3,4)$ of the Josephson potential change with $\theta/\pi$, the red dashed line is proportional to $\cos2\theta$, the inset shows an enlargement near $\theta/\pi=0.25$. (a) and (b) are calculated at $t_c=0.62$. (c) The ratio of the coefficient $E_J^{(1)}/E_J^{(2)}$ changes with $t_c$ with $\theta/\pi$ ranging from 0.25 to 0.3 (lines with $E_J^{(1)}/E_J^{(2)}>0$) and 0.25 to 0.2 (lines with $E_J^{(1)}/E_J^{(2)}<0$), the inset shows the coefficients change with $t_c$ at $\theta/\pi=0.25,0.26$. (d) The minimal point $\phi_{\rm{min}}/\pi$ changes with $\theta$ and interlayer couplings $t_c$.
  }
\label{sd-p}
\end{figure}

where the two terms correspond to single Cooper pair (two electrons) and double Cooper pairs (four electrons) co-tunneling across the s-d interface. The coefficients can be calculated as $E_J^{(1)}\propto t_c^2\cos2\theta$, and $E_J^{(2)}\propto t_c^4$ in the weak-tunneling limit ($t_c<\Delta$) \cite{supp}, which correspond to the tunneling of two electrons and four electrons, respectively. To extend the analysis beyond the weak coupling limit, we perform numerical calculations for the planar s-d junction by constructing a lattice model \cite{supp}. At $\theta=\pi/4$, the sign change of the d-wave gap cancels the odd harmonic, $E_J^{(1)}=0$, and the Josephson potential is dominated by a positive $\cos2\phi$ term \cite{Ioffe1999, Xiang2022}, shown by the blue line in Fig.~\ref{sd-p}(a). With negligible charging energy, the phase locks at one of its minima, $\phi=\pi/2$, yielding spontaneous TRS breaking (we take $\phi=\pi/2$ for concreteness) \cite{Ioffe1999}. When $\theta$ deviates from $\pi/4$, the single Cooper pair tunneling is allowed, a $\cos \phi$ term appears whose sign depends on whether $\theta$ is above or below $\pi/4$, shifting the minimum in opposite directions (green and yellow lines in Fig.\ref{sd-p}(a)). A Fourier decomposition of the numerically obtained potential (Fig.~\ref{sd-p}(b)) shows that $E_J^{(1)}(\theta)$ varies as $\cos2\theta$ (red dashed line), while near $\theta=\pi/4$, the $\cos2\phi$ term dominates, and the higher harmonics (e.g., $\cos3\phi$, $\cos4\phi$) are negligible, consistent with the suppression of higher-order tunneling processes. Since the critical condition for the spontaneous TRS breaking of the Josephson potential Eq.~\eqref{sd-pot} is $r\equiv |E_J^{(1)}/E_J^{(2)}|<4$, for which the minimum lies within $(0,\pi)$ \cite{Tummuru2022,Volkov2024,Guo2025}. We further evaluate the ratio $r$ as a function of $\theta$, with the critical value $E_J^{(1)}/E_J^{(2)}=4$ for the spontaneous TRS breaking, shown in Fig.~\ref{sd-p}(c). The ratio is symmetric about $\theta=\pi/4$, corresponding to $\Delta_{ds}=0$ in the $d$-wave pairing function, and follows $r\propto \cos2\theta/t_c^2$, which increases with $|\cos2\theta|$ but decreases with the s-d interlayer coupling. Thus, stronger s-d coupling stabilizes TRS-broken minima at $\phi\neq0,\pi$ over a wider range of $\theta$ away from $\pi/4$. Equivalently, the larger the s-d interface coupling $t_c$, the larger the permissible deviation of $\theta$ from $\pi/4$ for which spontaneous TRS breaking persists. This trend is corroborated in Fig.\ref{sd-p}(d), where the location of the free-energy minimum departs from $\pi/2$ as $\theta$ moves away from $\pi/4$, but stronger interlayer coupling increasingly pins the minimum near $\phi\approx \pi/2$, thereby stabilizing spontaneous TRS breaking.

\begin{figure}[!htbp]
	\centering
	\includegraphics[width=1\columnwidth]{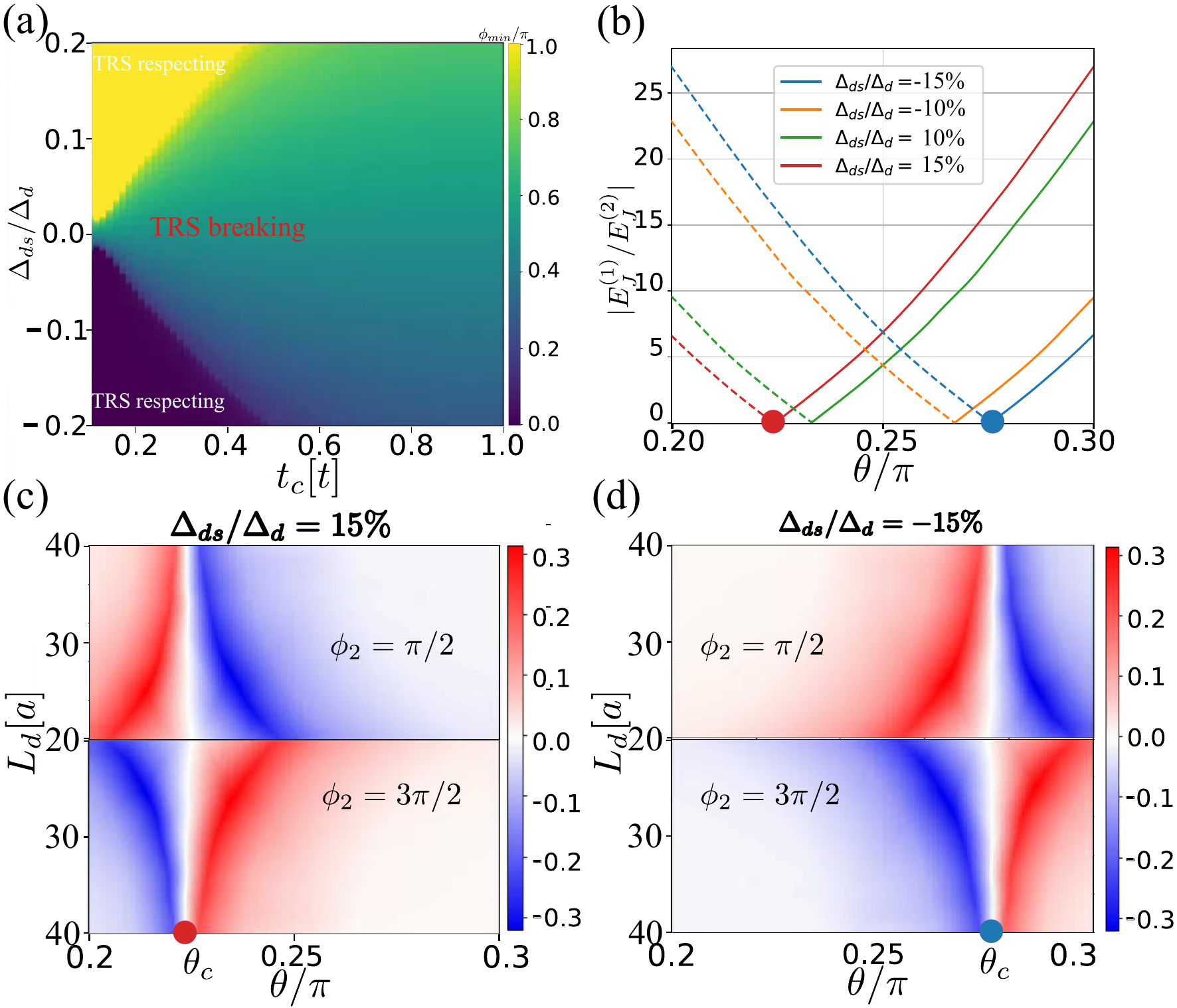}
	\caption{ For the s-d junction (a) $\theta=\pi/4$, the minimal point $\phi_{min}$ changes with $\Delta_{ds}/\Delta_d$; (b) The ratio of coefficient $|E_{J}^{(1)}/E_{J}^{(2)}|$ changes with $\theta$ with $\Delta_{ds}/\Delta_d=\pm10\%,\pm15\%$, the dots correspond to the dots in (c) and (d), dashed lines correspond negative values. For the s-d-s junction, the diode efficiency changes with $\theta$ and $L_d$ with $\Delta_{ds}/\Delta_d=-15\%$ (c), $15\%$ (d), the calculation is the same as that in Fig.~\ref{sds-jde}(e).
  }
\label{sds-bc4}
\end{figure}

{\it JDE with nonzero $\Delta_{ds}$ in $d$-wave superconductor.$-$} In the planar s-d-s junction system, single Cooper pair tunneling is essential for generating the JDE and can be activated either by rotating the lobe angle away from $\theta = \pi/4$ or by introducing an isotropic component in the d-wave superconductor, quantified by a nonzero $\Delta_{ds}$ with $\Delta_{ds}/\Delta_d$ quantifying the s-wave admixture  \cite{Sato2017,Zhu2021,Zhu2023}. In the latter scenario, the single Cooper pair tunneling remains allowed even at $\theta=\pi/4$, and the Josephson potential retains the form of Eq.~\eqref{sd-pot}, with the coefficients that now depend on $\Delta_{ds}/\Delta_d$, rather than $\theta$. Then, we calculate the potential minimal point changes with $\Delta_{ds}/\Delta_d$ and $t_c$, shown in Fig.~\ref{sds-bc4}(a). Because the $\cos \phi$ and $\cos 2\phi$ terms originate from second- and fourth-order electron tunneling processes, respectively, the earlier conclusion that larger interface coupling suppresses shifts of the free-energy minimum remains. Consequently, with asymmetric interface couplings, the spontaneous TRS breaking remains, and there exists a finite JDE even at $\theta=\pi/4$ for the nonzero $\Delta_{ds}/\Delta_d$. Furthermore, as $\theta$ deviates from $\pi/4$ further, the $\cos\phi_1$ term contribution can be driven through zero and change sign if the effects of $\Delta_{ds}$ and of $\theta-\pi/4$ give the opposite sign of the $\cos\phi_1$ term (Fig.~\ref{sds-bc4}(b)). Thus, the critical angle $\theta_c$ at which the JDE reverses polarity (also zero JDE) is shifted above or below $\pi/4$, depending on $\Delta_{ds}/\Delta_d>(<)0$ (Fig.~\ref{sds-bc4}(c)(d)). Equivalently, a nonzero JDE at $\theta=\pi/4$ directly signals nonzero $\Delta_{ds}/\Delta_d$, and the polarity of $\eta$ tracks the sign of the sign of $\Delta_{ds}/\Delta_d$.

{\it Discussion and summary} In this junction, the TRS breaking is achieved via a spontaneous $\pm\pi/2$ phase difference across the strongly coupled s-d interface without a magnetic field. Experimentally, such asymmetry can be engineered by inserting a tunnel barrier at one interface while keeping the other transparent or by deliberately creating a large contrast in contact area to enhance the associated Josephson coupling. In addition, the effective Josephson coupling between the two $s$‑wave electrodes can be tuned by introducing a direct weak link between them  \cite{supp}.

In summary, we propose the realization of JDE in a planar s-d-s junction as a probe of pairing symmetry in high-Tc superconductors. The polarity and magnitude of diode efficiency are tunable by adjusting the crystallographic angle of the d-wave superconductor or by introducing a breaking of its native $C_4$ pairing symmetry. Our work establishes a fundamental connection between the JDE in high-$T_c$ superconductors and the underlying pairing function, proposing non-reciprocal transport as a direct probe of pairing symmetry in high-Tc superconductors.

\begin{acknowledgments}
We acknowledge useful discussions with Noah F. Q. Yuan, Qiang-Qiang Gu, Xiang-Yu Zeng and Ren-Jie Zhang. X. Liu acknowledges the support of the Innovation Program for Quantum Science and Technology (Grant No. 2021ZD0302700), the National Natural Science Foundation of China (NSFC) (Grant No. 12074133), the Shanghai Science and Technology Innovation Action Plan (Grant  No. 24LZ1400800) and the Project supported by Cultivation Project of Shanghai Research Center for Quantum Sciences (Grant No. LZPY2024).
\end{acknowledgments}

\bibliography{Pl-sds-JDE-main-ref}

@Article{Scaff1947,
  author  = {Scaff, J. H. and Ohl, R. S.},
  journal = {The Bell System Technical Journal},
  title   = {Development of silicon crystal rectifiers for microwave radar receivers},
  year    = {1947},
  number  = {1},
  pages   = {1-30},
  volume  = {26},
  doi     = {10.1002/j.1538-7305.1947.tb01310.x},
}

@Article{Shockley1949,
  author   = {Shockley, W.},
  journal  = {Bell System Technical Journal},
  title    = {The Theory of p-n Junctions in Semiconductors and p-n Junction Transistors},
  year     = {1949},
  number   = {3},
  pages    = {435-489},
  volume   = {28},
  doi      = {https://doi.org/10.1002/j.1538-7305.1949.tb03645.x},
  url      = {https://onlinelibrary.wiley.com/doi/abs/10.1002/j.1538-7305.1949.tb03645.x},
}

@Article{Jiang2022,
  author   = {Jiang, Kun and Hu, Jiangping},
  journal  = {Nature Physics},
  title    = {Superconducting diode effects},
  year     = {2022},
  issn     = {1745-2481},
  number   = {10},
  pages    = {1145--1146},
  volume   = {18},
  doi      = {10.1038/s41567-022-01701-0},
  refid    = {Jiang2022},
  url      = {https://doi.org/10.1038/s41567-022-01701-0},
}

@Article{Zhang2022,
  author    = {Zhang, Yi and Gu, Yuhao and Li, Pengfei and Hu, Jiangping and Jiang, Kun},
  journal   = {Phys. Rev. X},
  title     = {General Theory of Josephson Diodes},
  year      = {2022},
  month     = {Nov},
  pages     = {041013},
  volume    = {12},
  doi       = {10.1103/PhysRevX.12.041013},
  issue     = {4},
  numpages  = {11},
  publisher = {American Physical Society},
  url       = {https://link.aps.org/doi/10.1103/PhysRevX.12.041013},
}

@Article{Narita2022,
  author   = {Narita, Hideki and Ishizuka, Jun and Kawarazaki, Ryo and Kan, Daisuke and Shiota, Yoichi and Moriyama, Takahiro and Shimakawa, Yuichi and Ognev, Alexey V. and Samardak, Alexander S. and Yanase, Youichi and Ono, Teruo},
  journal  = {Nature Nanotechnology},
  title    = {Field-free superconducting diode effect in noncentrosymmetric superconductor/ferromagnet multilayers},
  year     = {2022},
  issn     = {1748-3395},
  number   = {8},
  pages    = {823--828},
  volume   = {17},
  doi      = {10.1038/s41565-022-01159-4},
  refid    = {Narita2022},
  url      = {https://doi.org/10.1038/s41565-022-01159-4},
}

@Article{Davydova2022,
  author   = {Margarita Davydova and Saranesh Prembabu and Liang Fu},
  journal  = {Science Advances},
  title    = {Universal Josephson diode effect},
  year     = {2022},
  number   = {23},
  pages    = {eabo0309},
  volume   = {8},
  doi      = {10.1126/sciadv.abo0309},
  url      = {https://www.science.org/doi/abs/10.1126/sciadv.abo0309},
}

@Article{Banerjee2023,
  author    = {Banerjee, Abhishek and Geier, Max and Rahman, Md Ahnaf and Thomas, Candice and Wang, Tian and Manfra, Michael J. and Flensberg, Karsten and Marcus, Charles M.},
  journal   = {Phys. Rev. Lett.},
  title     = {Phase Asymmetry of Andreev Spectra from Cooper-Pair Momentum},
  year      = {2023},
  month     = {Nov},
  pages     = {196301},
  volume    = {131},
  doi       = {10.1103/PhysRevLett.131.196301},
  issue     = {19},
  numpages  = {6},
  publisher = {American Physical Society},
  url       = {https://link.aps.org/doi/10.1103/PhysRevLett.131.196301},
}

@Article{Legg2023,
  author    = {Legg, Henry F. and Laubscher, Katharina and Loss, Daniel and Klinovaja, Jelena},
  journal   = {Phys. Rev. B},
  title     = {Parity-protected superconducting diode effect in topological Josephson junctions},
  year      = {2023},
  month     = {Dec},
  pages     = {214520},
  volume    = {108},
  doi       = {10.1103/PhysRevB.108.214520},
  issue     = {21},
  numpages  = {7},
  publisher = {American Physical Society},
  url       = {https://link.aps.org/doi/10.1103/PhysRevB.108.214520},
}

@Article{Maiani2023,
  author    = {Maiani, Andrea and Flensberg, Karsten and Leijnse, Martin and Schrade, Constantin and Vaitiek\ifmmode \dot{e}\else \.{e}\fi{}nas, Saulius and Seoane Souto, Rub\'en},
  journal   = {Phys. Rev. B},
  title     = {Nonsinusoidal current-phase relations in semiconductor--superconductor-- ferromagnetic insulator devices},
  year      = {2023},
  month     = {Jun},
  pages     = {245415},
  volume    = {107},
  doi       = {10.1103/PhysRevB.107.245415},
  issue     = {24},
  numpages  = {12},
  publisher = {American Physical Society},
  url       = {https://link.aps.org/doi/10.1103/PhysRevB.107.245415},
}

@Article{Pal2022,
  author   = {Pal, Banabir and Chakraborty, Anirban and Sivakumar, Pranava K. and Davydova, Margarita and Gopi, Ajesh K. and Pandeya, Avanindra K. and Krieger, Jonas A. and Zhang, Yang and Date, Mihir and Ju, Sailong and Yuan, Noah and Schröter, Niels B. M. and Fu, Liang and Parkin, Stuart S. P.},
  journal  = {Nature Physics},
  title    = {Josephson diode effect from Cooper pair momentum in a topological semimetal},
  year     = {2022},
  issn     = {1745-2481},
  number   = {10},
  pages    = {1228--1233},
  volume   = {18},
  doi      = {10.1038/s41567-022-01699-5},
  refid    = {Pal2022},
  url      = {https://doi.org/10.1038/s41567-022-01699-5},
}

@Article{Steiner2023,
  author    = {Steiner, Jacob F. and Melischek, Larissa and Trahms, Martina and Franke, Katharina J. and von Oppen, Felix},
  journal   = {Phys. Rev. Lett.},
  title     = {Diode Effects in Current-Biased Josephson Junctions},
  year      = {2023},
  month     = {Apr},
  pages     = {177002},
  volume    = {130},
  doi       = {10.1103/PhysRevLett.130.177002},
  issue     = {17},
  numpages  = {6},
  publisher = {American Physical Society},
  url       = {https://link.aps.org/doi/10.1103/PhysRevLett.130.177002},
}

@Article{Yuan2022,
  author   = {Noah F. Q. Yuan and Liang Fu},
  journal  = {Proceedings of the National Academy of Sciences},
  title    = {Supercurrent diode effect and finite-momentum superconductors},
  year     = {2022},
  number   = {15},
  pages    = {e2119548119},
  volume   = {119},
  doi      = {10.1073/pnas.2119548119},
  url      = {https://www.pnas.org/doi/abs/10.1073/pnas.2119548119},
}

@Article{Legg2022,
  author    = {Legg, Henry F. and Loss, Daniel and Klinovaja, Jelena},
  journal   = {Phys. Rev. B},
  title     = {Superconducting diode effect due to magnetochiral anisotropy in topological insulators and Rashba nanowires},
  year      = {2022},
  month     = {Sep},
  pages     = {104501},
  volume    = {106},
  doi       = {10.1103/PhysRevB.106.104501},
  issue     = {10},
  numpages  = {8},
  publisher = {American Physical Society},
  url       = {https://link.aps.org/doi/10.1103/PhysRevB.106.104501},
}

@Article{Hu2023,
  author    = {Hu, Jin-Xin and Sun, Zi-Ting and Xie, Ying-Ming and Law, K. T.},
  journal   = {Phys. Rev. Lett.},
  title     = {Josephson Diode Effect Induced by Valley Polarization in Twisted Bilayer Graphene},
  year      = {2023},
  month     = {Jun},
  pages     = {266003},
  volume    = {130},
  doi       = {10.1103/PhysRevLett.130.266003},
  issue     = {26},
  numpages  = {6},
  publisher = {American Physical Society},
  url       = {https://link.aps.org/doi/10.1103/PhysRevLett.130.266003},
}

@Article{Lin2022,
  author   = {Lin, Jiang-Xiazi and Siriviboon, Phum and Scammell, Harley D. and Liu, Song and Rhodes, Daniel and Watanabe, K. and Taniguchi, T. and Hone, James and Scheurer, Mathias S. and Li, J. I. A.},
  journal  = {Nature Physics},
  title    = {Zero-field superconducting diode effect in small-twist-angle trilayer graphene},
  year     = {2022},
  issn     = {1745-2481},
  number   = {10},
  pages    = {1221--1227},
  volume   = {18},
  doi      = {10.1038/s41567-022-01700-1},
  refid    = {Lin2022},
  url      = {https://doi.org/10.1038/s41567-022-01700-1},
}

@Article{Wu2022,
  author   = {Wu, Heng and Wang, Yaojia and Xu, Yuanfeng and Sivakumar, Pranava K. and Pasco, Chris and Filippozzi, Ulderico and Parkin, Stuart S. P. and Zeng, Yu-Jia and McQueen, Tyrel and Ali, Mazhar N.},
  journal  = {Nature},
  title    = {The field-free Josephson diode in a van der Waals heterostructure},
  year     = {2022},
  issn     = {1476-4687},
  number   = {7907},
  pages    = {653--656},
  volume   = {604},
  doi      = {10.1038/s41586-022-04504-8},
  refid    = {Wu2022},
  url      = {https://doi.org/10.1038/s41586-022-04504-8},
}

@Article{Lu2023,
  author    = {Lu, Bo and Ikegaya, Satoshi and Burset, Pablo and Tanaka, Yukio and Nagaosa, Naoto},
  journal   = {Phys. Rev. Lett.},
  title     = {Tunable Josephson Diode Effect on the Surface of Topological Insulators},
  year      = {2023},
  month     = {Aug},
  pages     = {096001},
  volume    = {131},
  doi       = {10.1103/PhysRevLett.131.096001},
  issue     = {9},
  numpages  = {7},
  publisher = {American Physical Society},
  url       = {https://link.aps.org/doi/10.1103/PhysRevLett.131.096001},
}

@Article{Ghosh2024,
  author   = {Ghosh, Sanat and Patil, Vilas and Basu, Amit and Kuldeep and Dutta, Achintya and Jangade, Digambar A. and Kulkarni, Ruta and Thamizhavel, A. and Steiner, Jacob F. and von Oppen, Felix and Deshmukh, Mandar M.},
  journal  = {Nature Materials},
  title    = {High-temperature Josephson diode},
  year     = {2024},
  issn     = {1476-4660},
  number   = {5},
  pages    = {612--618},
  volume   = {23},
  doi      = {10.1038/s41563-024-01804-4},
  refid    = {Ghosh2024},
  url      = {https://doi.org/10.1038/s41563-024-01804-4},
}

@Article{Volkov2024,
  author    = {Volkov, Pavel A. and Lantagne-Hurtubise, \'Etienne and Tummuru, Tarun and Plugge, Stephan and Pixley, J. H. and Franz, Marcel},
  journal   = {Phys. Rev. B},
  title     = {Josephson diode effects in twisted nodal superconductors},
  year      = {2024},
  month     = {Mar},
  pages     = {094518},
  volume    = {109},
  doi       = {10.1103/PhysRevB.109.094518},
  issue     = {9},
  numpages  = {14},
  publisher = {American Physical Society},
  url       = {https://link.aps.org/doi/10.1103/PhysRevB.109.094518},
}

@Article{Zhu2023,
  author    = {Zhu, Yuying and Wang, Heng and Wang, Zechao and Hu, Shuxu and Gu, Genda and Zhu, Jing and Zhang, Ding and Xue, Qi-Kun},
  journal   = {Phys. Rev. B},
  title     = {Persistent Josephson tunneling between ${\mathrm{Bi}}_{2}{\mathrm{Sr}}_{2}{\mathrm{CaCu}}_{2}{\mathrm{O}}_{8+x}$ flakes twisted by ${45}^{\ensuremath{\circ}}$ across the superconducting dome},
  year      = {2023},
  month     = {Nov},
  pages     = {174508},
  volume    = {108},
  doi       = {10.1103/PhysRevB.108.174508},
  issue     = {17},
  numpages  = {9},
  publisher = {American Physical Society},
  url       = {https://link.aps.org/doi/10.1103/PhysRevB.108.174508},
}

@Article{Zhu2021,
  author    = {Zhu, Yuying and Liao, Menghan and Zhang, Qinghua and Xie, Hong-Yi and Meng, Fanqi and Liu, Yaowu and Bai, Zhonghua and Ji, Shuaihua and Zhang, Jin and Jiang, Kaili and Zhong, Ruidan and Schneeloch, John and Gu, Genda and Gu, Lin and Ma, Xucun and Zhang, Ding and Xue, Qi-Kun},
  journal   = {Phys. Rev. X},
  title     = {Presence of $s$-Wave Pairing in Josephson Junctions Made of Twisted Ultrathin ${\mathrm{Bi}}_{2}{\mathrm{Sr}}_{2}{\mathrm{CaCu}}_{2}{\mathrm{O}}_{8+x}$ Flakes},
  year      = {2021},
  month     = {Jul},
  pages     = {031011},
  volume    = {11},
  doi       = {10.1103/PhysRevX.11.031011},
  issue     = {3},
  numpages  = {13},
  publisher = {American Physical Society},
  url       = {https://link.aps.org/doi/10.1103/PhysRevX.11.031011},
}

@Article{Tummuru2022,
  author    = {Tummuru, Tarun and Plugge, Stephan and Franz, Marcel},
  journal   = {Phys. Rev. B},
  title     = {Josephson effects in twisted cuprate bilayers},
  year      = {2022},
  month     = {Feb},
  pages     = {064501},
  volume    = {105},
  doi       = {10.1103/PhysRevB.105.064501},
  issue     = {6},
  numpages  = {14},
  publisher = {American Physical Society},
  url       = {https://link.aps.org/doi/10.1103/PhysRevB.105.064501},
}

@Article{Zhao2023,
  author   = {S. Y. Frank Zhao and Xiaomeng Cui and Pavel A. Volkov and Hyobin Yoo and Sangmin Lee and Jules A. Gardener and Austin J. Akey and Rebecca Engelke and Yuval Ronen and Ruidan Zhong and Genda Gu and Stephan Plugge and Tarun Tummuru and Miyoung Kim and Marcel Franz and Jedediah H. Pixley and Nicola Poccia and Philip Kim},
  journal  = {Science},
  title    = {Time-reversal symmetry breaking superconductivity between twisted cuprate superconductors},
  year     = {2023},
  number   = {6677},
  pages    = {1422-1427},
  volume   = {382},
  doi      = {10.1126/science.abl8371},
  url      = {https://www.science.org/doi/abs/10.1126/science.abl8371},
}

@Article{Patel2024,
  author    = {Patel, Hrishikesh and Pathak, Vedangi and Can, Oguzhan and Potter, Andrew C. and Franz, Marcel},
  journal   = {Phys. Rev. Lett.},
  title     = {$d$-Mon: A Transmon with Strong Anharmonicity Based on Planar $c$-Axis Tunneling Junction between $d$-Wave and $s$-Wave Superconductors},
  year      = {2024},
  month     = {Jan},
  pages     = {017002},
  volume    = {132},
  doi       = {10.1103/PhysRevLett.132.017002},
  issue     = {1},
  numpages  = {6},
  publisher = {American Physical Society},
  url       = {https://link.aps.org/doi/10.1103/PhysRevLett.132.017002},
}

@Article{Sato2017,
  author   = {Sato, Y. and Kasahara, S. and Murayama, H. and Kasahara, Y. and Moon, E.-G. and Nishizaki, T. and Loew, T. and Porras, J. and Keimer, B. and Shibauchi, T. and Matsuda, Y.},
  journal  = {Nature Physics},
  title    = {Thermodynamic evidence for a nematic phase transition at the onset of the pseudogap in YBa2Cu3Oy},
  year     = {2017},
  issn     = {1745-2481},
  number   = {11},
  pages    = {1074--1078},
  volume   = {13},
  doi      = {10.1038/nphys4205},
  refid    = {Sato2017},
  url      = {https://doi.org/10.1038/nphys4205},
}

@Unpublished{supp,
  note = {See supplemental materials for:},
}

@Article{Wang2023,
  author   = {Wang, Heng and Zhu, Yuying and Bai, Zhonghua and Wang, Zechao and Hu, Shuxu and Xie, Hong-Yi and Hu, Xiaopeng and Cui, Jian and Huang, Miaoling and Chen, Jianhao and Ding, Ying and Zhao, Lin and Li, Xinyan and Zhang, Qinghua and Gu, Lin and Zhou, X. J. and Zhu, Jing and Zhang, Ding and Xue, Qi-Kun},
  journal  = {Nature Communications},
  title    = {Prominent Josephson tunneling between twisted single copper oxide planes of Bi2Sr2-xLaxCuO6+y},
  year     = {2023},
  issn     = {2041-1723},
  number   = {1},
  pages    = {5201},
  volume   = {14},
  doi      = {10.1038/s41467-023-40525-1},
  refid    = {Wang2023},
  url      = {https://doi.org/10.1038/s41467-023-40525-1},
}

@Article{Liao2018,
  author    = {Liao, Menghan and Zhu, Yuying and Zhang, Jin and Zhong, Ruidan and Schneeloch, John and Gu, Genda and Jiang, Kaili and Zhang, Ding and Ma, Xucun and Xue, Qi-Kun},
  journal   = {Nano Lett.},
  title     = {Superconductor-Insulator Transitions in Exfoliated Bi2Sr2CaCu2O8+δ Flakes},
  year      = {2018},
  issn      = {1530-6984},
  month     = sep,
  number    = {9},
  pages     = {5660--5665},
  volume    = {18},
  comment   = {doi: 10.1021/acs.nanolett.8b02183},
  doi       = {10.1021/acs.nanolett.8b02183},
  publisher = {American Chemical Society},
  url       = {https://doi.org/10.1021/acs.nanolett.8b02183},
}

@Article{Nadeem2023,
  author   = {Nadeem, Muhammad and Fuhrer, Michael S. and Wang, Xiaolin},
  journal  = {Nature Reviews Physics},
  title    = {The superconducting diode effect},
  year     = {2023},
  issn     = {2522-5820},
  number   = {10},
  pages    = {558--577},
  volume   = {5},
  doi      = {10.1038/s42254-023-00632-w},
  refid    = {Nadeem2023},
  url      = {https://doi.org/10.1038/s42254-023-00632-w},
}

@Article{Reinhardt2024,
  author   = {Reinhardt, S. and Ascherl, T. and Costa, A. and Berger, J. and Gronin, S. and Gardner, G. C. and Lindemann, T. and Manfra, M. J. and Fabian, J. and Kochan, D. and Strunk, C. and Paradiso, N.},
  journal  = {Nature Communications},
  title    = {Link between supercurrent diode and anomalous Josephson effect revealed by gate-controlled interferometry},
  year     = {2024},
  issn     = {2041-1723},
  number   = {1},
  pages    = {4413},
  volume   = {15},
  doi      = {10.1038/s41467-024-48741-z},
  refid    = {Reinhardt2024},
  url      = {https://doi.org/10.1038/s41467-024-48741-z},
}

@Book{Xiang2022,
  author    = {Xiang, Tao and Wu, Congjun},
  publisher = {Cambridge University Press},
  title     = {D-wave Superconductivity},
  year      = {2022},
  place     = {Cambridge},
}

@Article{Guo2025,
  author    = {Guo, Guo-Liang and Pan, Xiao-Hong and Liu, Xin},
  journal   = {Phys. Rev. B},
  title     = {${\ensuremath{\phi}}_{0}$ junction and Josephson diode effect in high-temperature superconductors},
  year      = {2025},
  month     = {Jul},
  pages     = {014509},
  volume    = {112},
  doi       = {10.1103/PhysRevB.112.014509},
  issue     = {1},
  numpages  = {21},
  publisher = {American Physical Society},
  url       = {https://link.aps.org/doi/10.1103/PhysRevB.112.014509},
}

@Article{Ioffe1999,
  author   = {Ioffe, Lev B. and Geshkenbein, Vadim B. and Feigel'man, Mikhail V. and Fauchère, Alban L. and Blatter, Gianni},
  journal  = {Nature},
  title    = {Environmentally decoupled sds -wave Josephson junctions for quantum computing},
  year     = {1999},
  issn     = {1476-4687},
  number   = {6729},
  pages    = {679--681},
  volume   = {398},
  doi      = {10.1038/19464},
  refid    = {Ioffe1999},
  url      = {https://doi.org/10.1038/19464},
}

@Article{Zheng2025,
  author        = {{Zheng}, Wayne and {Cheng}, Tao and {Yue}, Zheng-Yuan and {Zhang}, Fu-Chun and {Chen}, Wei-Qiang and {Gu}, Zheng-Cheng},
  journal       = {arXiv e-prints},
  title         = {{Competing $s$-wave pairing in overdoped $t$-$J$ model}},
  year          = {2025},
  month         = sep,
  pages         = {arXiv:2509.22473},
  adsurl        = {https://ui.adsabs.harvard.edu/abs/2025arXiv250922473Z},
  archiveprefix = {arXiv},
  doi           = {10.48550/arXiv.2509.22473},
  eid           = {arXiv:2509.22473},
  eprint        = {2509.22473},
  primaryclass  = {cond-mat.str-el},
}

@Article{Tsuei2000,
  author    = {Tsuei, C. C. and Kirtley, J. R.},
  journal   = {Rev. Mod. Phys.},
  title     = {Pairing symmetry in cuprate superconductors},
  year      = {2000},
  month     = {Oct},
  pages     = {969--1016},
  volume    = {72},
  doi       = {10.1103/RevModPhys.72.969},
  issue     = {4},
  numpages  = {0},
  publisher = {American Physical Society},
  url       = {https://link.aps.org/doi/10.1103/RevModPhys.72.969},
}

@Article{Mei2025,
  author        = {{Mei}, Jiong and {Qin}, Shengshan and {Hu}, Jiangping},
  journal       = {arXiv e-prints},
  title         = {{Interband-Pairing-Boosted Supercurrent Diode Effect in Multiband Superconductors}},
  year          = {2025},
  month         = oct,
  pages         = {arXiv:2510.15788},
  adsurl        = {https://ui.adsabs.harvard.edu/abs/2025arXiv251015788M},
  archiveprefix = {arXiv},
  doi           = {10.48550/arXiv.2510.15788},
  eid           = {arXiv:2510.15788},
  eprint        = {2510.15788},
  primaryclass  = {cond-mat.supr-con},
}

@Article{Kochan2023,
  author        = {{Kochan}, Denis and {Costa}, Andreas and {Zhumagulov}, Iaroslav and {{\v{Z}}uti{\'c}}, Igor},
  journal       = {arXiv e-prints},
  title         = {{Phenomenological Theory of the Supercurrent Diode Effect: The Lifshitz Invariant}},
  year          = {2023},
  month         = mar,
  pages         = {arXiv:2303.11975},
  adsurl        = {https://ui.adsabs.harvard.edu/abs/2023arXiv230311975K},
  archiveprefix = {arXiv},
  doi           = {10.48550/arXiv.2303.11975},
  eid           = {arXiv:2303.11975},
  eprint        = {2303.11975},
  primaryclass  = {cond-mat.supr-con},
}

@Article{Wang2025,
  author        = {{Wang}, Bao-Zong and {Li}, Zi-Kai and {Li}, Zhong-Da and {Liu}, Xiong-Jun},
  journal       = {arXiv e-prints},
  title         = {{Giant and robust Josephson diode effect in multiband topological nanowires}},
  year          = {2025},
  month         = oct,
  pages         = {arXiv:2510.05772},
  adsurl        = {https://ui.adsabs.harvard.edu/abs/2025arXiv251005772W},
  archiveprefix = {arXiv},
  doi           = {10.48550/arXiv.2510.05772},
  eid           = {arXiv:2510.05772},
  eprint        = {2510.05772},
  primaryclass  = {cond-mat.mes-hall},
}

@Article{Fradkin2015,
  author    = {Fradkin, Eduardo and Kivelson, Steven A. and Tranquada, John M.},
  journal   = {Rev. Mod. Phys.},
  title     = {Colloquium: Theory of intertwined orders in high temperature superconductors},
  year      = {2015},
  month     = {May},
  pages     = {457--482},
  volume    = {87},
  doi       = {10.1103/RevModPhys.87.457},
  issue     = {2},
  numpages  = {26},
  publisher = {American Physical Society},
  url       = {https://link.aps.org/doi/10.1103/RevModPhys.87.457},
}

@Article{Leng2012,
  author    = {Leng, Xiang and Garcia-Barriocanal, Javier and Yang, Boyi and Lee, Yeonbae and Kinney, J. and Goldman, A. M.},
  journal   = {Phys. Rev. Lett.},
  title     = {Indications of an Electronic Phase Transition in Two-Dimensional Superconducting ${\mathrm{YBa}}_{2}{\mathrm{Cu}}_{3}{\mathrm{O}}_{7\ensuremath{-}x}$ Thin Films Induced by Electrostatic Doping},
  year      = {2012},
  month     = {Feb},
  pages     = {067004},
  volume    = {108},
  doi       = {10.1103/PhysRevLett.108.067004},
  issue     = {6},
  numpages  = {5},
  publisher = {American Physical Society},
  url       = {https://link.aps.org/doi/10.1103/PhysRevLett.108.067004},
}

@Article{Cai2016,
  author   = {Cai, Peng and Ruan, Wei and Peng, Yingying and Ye, Cun and Li, Xintong and Hao, Zhenqi and Zhou, Xingjiang and Lee, Dung-Hai and Wang, Yayu},
  journal  = {Nature Physics},
  title    = {Visualizing the evolution from the Mott insulator to a charge-ordered insulator in lightly doped cuprates},
  year     = {2016},
  issn     = {1745-2481},
  number   = {11},
  pages    = {1047--1051},
  volume   = {12},
  doi      = {10.1038/nphys3840},
  refid    = {Cai2016},
  url      = {https://doi.org/10.1038/nphys3840},
}

@Article{Jacobs2016,
  author    = {Jacobs, Th. and Simsek, Y. and Koval, Y. and M\"uller, P. and Krasnov, V. M.},
  journal   = {Phys. Rev. Lett.},
  title     = {Sequence of Quantum Phase Transitions in ${\mathrm{Bi}}_{2}{\mathrm{Sr}}_{2}{\mathrm{CaCu}}_{2}{\mathrm{O}}_{8+\ensuremath{\delta}}$ Cuprates Revealed by In Situ Electrical Doping of One and the Same Sample},
  year      = {2016},
  month     = {Feb},
  pages     = {067001},
  volume    = {116},
  doi       = {10.1103/PhysRevLett.116.067001},
  issue     = {6},
  numpages  = {6},
  publisher = {American Physical Society},
  url       = {https://link.aps.org/doi/10.1103/PhysRevLett.116.067001},
}

@Article{McElroy2005,
  author    = {McElroy, K. and Lee, D.-H. and Hoffman, J. E. and Lang, K. M. and Lee, J. and Hudson, E. W. and Eisaki, H. and Uchida, S. and Davis, J. C.},
  journal   = {Phys. Rev. Lett.},
  title     = {Coincidence of Checkerboard Charge Order and Antinodal State Decoherence in Strongly Underdoped Superconducting ${\mathrm{Bi}}_{2}{\mathrm{Sr}}_{2}\mathrm{Ca}{\mathrm{Cu}}_{2}{\mathrm{O}}_{8+\ensuremath{\delta}}$},
  year      = {2005},
  month     = {May},
  pages     = {197005},
  volume    = {94},
  doi       = {10.1103/PhysRevLett.94.197005},
  issue     = {19},
  numpages  = {4},
  publisher = {American Physical Society},
  url       = {https://link.aps.org/doi/10.1103/PhysRevLett.94.197005},
}

@Article{Hirschfeld1993,
  author    = {Hirschfeld, Peter J. and Goldenfeld, Nigel},
  journal   = {Phys. Rev. B},
  title     = {Effect of strong scattering on the low-temperature penetration depth of a d-wave superconductor},
  year      = {1993},
  month     = {Aug},
  pages     = {4219--4222},
  volume    = {48},
  doi       = {10.1103/PhysRevB.48.4219},
  issue     = {6},
  numpages  = {0},
  publisher = {American Physical Society},
  url       = {https://link.aps.org/doi/10.1103/PhysRevB.48.4219},
}

@Article{Wang2025a,
  author        = {{Wang}, Heng and {Zhu}, Yuying and {Bai}, Zhonghua and {Lyu}, Zhaozheng and {Yang}, Jiangang and {Zhao}, Lin and {Zhou}, X.~J. and {Gu}, Genda and {Xue}, Qi-Kun and {Zhang}, Ding},
  journal       = {arXiv e-prints},
  title         = {{Quantum superconducting diode effect with perfect efficiency above liquid-nitrogen temperature}},
  year          = {2025},
  month         = sep,
  pages         = {arXiv:2509.24764},
  adsurl        = {https://ui.adsabs.harvard.edu/abs/2025arXiv250924764W},
  archiveprefix = {arXiv},
  doi           = {10.48550/arXiv.2509.24764},
  eid           = {arXiv:2509.24764},
  eprint        = {2509.24764},
  primaryclass  = {cond-mat.supr-con},
}

\end{document}